**Invasive species, extreme fire risk, and toxin release under a changing climate**

K.R. Miner[1,2], L. A. Meyerson[2]
1. Climate Change Institute and School of Earth and Climate Sciences, University of Maine, Orono, ME 04469
2. Department of Natural Resources Science, University of Rhode Island, Kingston, RI 02881

**Abstract:**
Mediterranean ecosystems such as those found in California, Central Chile, Southern Europe, and Southwest Australia host numerous, diverse, fire-adapted micro-ecosystems. These micro-ecosystems are as diverse as mountainous conifer to desert-like chaparral communities. Over the last few centuries, human intervention, invasive species, and climate warming have drastically affected the composition and health of Mediterranean ecosystems on almost every continent.  Increased fuel load from fire suppression policies and the continued range expansion of non-native insects and plants, some driven by long-term drought, produced the deadliest wildfire season on record in 2018. As a consequence of these fires, a large number of structures are destroyed, releasing household chemicals into the environment as uncontrolled toxins. The mobilization of these materials can lead to health risks and disruption in both human and natural systems. This article identifies drivers that led to a structural weakening of the mosaic of fire-adapted ecosystems in California, and subsequently increased the risk of destructive and explosive wildfires throughout the state. Under a new climate regime, managing the impacts on systems moving out-of-phase with natural processes may protect lives and ensure the stability of ecosystem services.

**Introduction**

California's Mediterranean climate and fire-adapted ecosystems, like those across the world, host a wide diversity of plant and animal species that require frequent fires to thrive. For example, fire facilitates the release and dispersal of seeds by some members of the *Pinophyta* (Conifer) family, or the removal of overgrown brush on the forest floor, leaving space for the growth of first generational ground cover plants, seedlings and other serotinous species (e.g., Kilgore and Taylor 1979; He et al. 2015). This fire-directed ecosystem structure was maintained for many generations in California by indigenous communities, including the Karok and Yurok, known for integrating fire into their land management practices (Denevan 1992; Keeley 2002; Little 2018). However, with the arrival of European development and infrastructure in Mediterranean climates throughout the world, fire suppression became the governing philosophy (Parsons and Debenedetti 1979). For the last few decades, staunch fire suppression policies implemented to protect built property have minimized the number of fires across the California landscapes, thereby increasing the fuel load and changing the natural cycling of fire-reliant species (Parsons and Debenedetti 1979; Collins et al. 2019).  Combined with drought conditions and the expanded range of flammable invasive species, these fire suppression policies have primed California ecosystems for exceedingly hot and destructive fires (Diffenbaugh et al. 2015; Steel et al. 2015).

With an overlap between human development and fire-prone natural systems, hotter fires can lead to the unregulated release of household and industrial products from burnt houses, businesses, and production facilities. These compounds can be extremely toxic and include heating oil and tar, cleaning products, insulation and flame-retardants, asbestos, plastics, PCBs and benzene. This little-understood impact of the increased heat and range of wildfires can cause significant health impacts on humans and natural systems, with effects that can span generations.

The goal of this article, therefore, is to investigate the cyclical system dynamics between invasive species, climate change, increased fire range, and the release of toxic chemicals within Mediterranean climates. Understanding the interplay between invasive species, ecosystem-destabilizing chemicals, and temperature increases will allow local authorities to anticipate and better plan for the danger of extreme wildfire risk in Mediterranean ecosystems across the globe.

**Fire dynamics in Mediterranean ecosystems**

Within multiple California ecosystems, the introduction and range expansion of invasive flora produce biomass that acts as surplus fuel during wildfires (Keeley 2001; Rao and Allen 2010; Bell et al. 2017; Calvino-Cancela et al. 2017). Invasive plants such as members of the Tamarix family, including *Tamarix ramosissima* (salt cedar) and the Myrtaceae family, with over 60 species of *Eucalyptus,* thrive in riparian environments. *Avena fatua* (wild oats), *Lolium perenne (*ryegrass) and *Arundo donax* (giant reed grass) have invaded grasslands with flammable biomass, that can be faster than native species to recover after a fire due to a persistent seed bank and rapid growth rates (Bell et al. 2017) (e.g., Keeley 2001; California Native Plant Society 2010). Fire-prone eucalyptus trees can further exacerbate forest wildfires, as they have during the 2017-2019 Australian fire seasons, spreading the flames to great distances with explosive bark and sap (Bell et al. 2017; Calvino-Cancela et al. 2017). These invasive plants increase the intensity and range of wildfires, impacting the survivability of co-located native plants (Bell et al. 2017).

In addition to floral invasives, over 200 native species of the fauna species *Scolytidae* (including bark beetles), expand in number and range as drought and dense forests allow them to quickly attack trees weakened by the resource drain (Bentz et al. 2010; Bell et al. 2017). The significant damage left by the bark beetle includes predation and death of old-growth pines, disproportionately affecting these targeted species. *Scolytidae* species are critical for maintaining healthy forest structure under normal conditions but may kill excessively during drought, laying eggs under the bark of vulnerable trees (Bentz et al. 2010). Bark beetles also suffer higher mortality under cold temperatures, which prevented the population from expanding unsustainably in the past. Still, temperature increase from climate change has allowed the beetles to expand their range from Mexico to Alaska (Bentz et al. 2010). As forests decrease in size and strength, they absorb less carbon dioxide, leading to further heating and deforestation. The additional dry material from dead trees can fuel a wildfire able to move up trees and across treetops, in a movement known as crowning, jumping over fire blocks to burn large areas (Steel et al. 2015). These scorching fires can also remove or damage topsoil quality and nutrient supply, decreasing the chance of native and serotinous plant regrowth  (Knelman et al. 2015).

As climate change continues to warm and dry Mediterranean ecosystems, the consequences are widespread (Flato et al. 2013; Kloos et al. 2013; Roe et al. 2016; Schwartz 2018). Warming systems have led to multi-year snowpack lows in California's Sierra Nevada range (Belmecheri et al. 2016) with co-occurring high temperatures and drought conditions expected to continue seasonally (Asner et al. 2016). Combined with a low precipitation rate that decreases soil hydration, the potential for hotter, wide-reaching wildfires escalates (Asner et al. 2016). In addition to the loss of snowpack and precipitation variability, drought introduces even greater vulnerability to insect attack for many trees and plant species. Forest fires burn at different temperatures, ranging from 800 °C to 1200 °C, depending on the fuel and oxygen availability. As the number of dead trees and plant material increases, the opportunity for significant dry fuel loading also grows, amplifying the risk of disastrous fires (Kilgore and Taylor 1979; Parsons and Debenedetti 1979; He et al. 2015; Steel et al. 2015). With over 129 million dead trees in the state of California from bark beetles in recent years (Fire 2017), both the forest fuel-load and the opportunity for invasive flora range advance has increased exponentially.

**Household aids become toxins in the environment**

Human development within natural systems in California, combined with drought-like conditions and fuel loading driven by invasive species, increases the risk of an uncontrolled release of human-made chemicals during wildfires. The discharge of household and industrial chemicals during structural burns in fires, in effect, transform useful household aids into toxins. When a home in a forest burns, the materials that compose the building structure, any fuel sources or stored petroleum, sewage and septic systems, cleaning products, and every-day use objects are transformed into ash or carried directly in water into the environment (Johnston 2018). Structural ash may contain an amalgamate of chemical particulates, including cancer-causing dibenzo-p-dioxin (PCDD), PFAs, PAHs, benzene, organochlorines, bleach and cleaning compounds, petroleum byproducts, naphthalene, carbon monoxide, and airborne asbestos (Fernández-Fernández et al. 2015; Kibet et al. 2017). The compounds that make up these products can range from harmful to extremely dangerous to health, known to contribute to cancer, diabetes, congenital disabilities and obesity in humans (Ruzzin 2012; Lee et al. 2014; Denise K. Reaves, Erika Ginsburg, John J. Bang 2015; Ngwa et al. 2015; Miner et al. 2018b). Compounds found in these structures can be concentrated in the fire ash, leading to an indiscriminate release of regulated chemicals into the near environment. Once released, they move through the atmosphere and watershed in unmeasurable quantities, where they permeate and damage natural ecosystems (Lafrenière et al. 2006; Miner et al. 2017; Mackay et al. 2018). Additional toxin distribution through atmospheric transport and re-deposition may occur during precipitation, further expanding the range to impact both regional and international ecosystems (Hansen et al. 2015; Friedman and Selin 2016; Miner et al. 2018a). Precipitation following burn events may also transport chemical particulates throughout the watershed and into streams and the ocean, dangerously impacting fish stock and marine life (Kibet et al. 2017; Miner et al. 2018c; Noestheden et al. 2018). The flux of household chemicals into local ecosystems increases the chance of biological uptake and accumulation, as well as the residence time and background concentration of the chemicals (Daly and Wania 2005; Bizzotto et al. 2009; Walters et al. 2016).

Any uncontrolled flux of known toxins into the environment at unknown rates has the potential for significant consequences for natural and human life within the ecosystem (Bogdal et al. 2009; Lee et al. 2014; Ren et al. 2016; Wang et al. 2016; Mackay et al. 2018). System instability predicated by insect or vegetation die-off or inter-generational bioaccumulation can exacerbate vulnerabilities in an already drought-prone system (Blais et al. 2003; Elliott et al. 2012; Diffenbaugh et al. 2015; Villa et al. 2017). Structural ecosystem weaknesses from drought and temperature fluctuations are exploited by invasive species- a dynamic that could increase with further damage from chemical uptake (Figure 1). Significantly, as chemicals released into the environment can destabilize a natural structure or cause system die-off, large quantities of chemicals released in a wildfire can damage the ecosystem for generations. The combination of built environment expansion, fire suppression policies, drought, and fuel loading from invasive species increases the opportunity for direct chemical release during wildfires.

**The potential for cascading system destabilization**

Cascading system failure describes the destruction of foundational aspects of a system, such as necessary prey or vital habitat, leading to reverberating damage throughout the system structure. As chemicals increase in concentration in the environment, they can lead to non-linear health impacts or largescale die-offs in both flora and fauna (EPA 2016; Reid et al. 2016; Kibet et al. 2017). The transport of foreign chemical species into ecosystems during wildfires can destabilize trophic-level predation, changing ecosystem structure, and resilience (Walters et al. 2016; Mackay et al. 2018). The combination of drought or unseasonal flooding and bioaccumulation of toxins can decrease the survival rate of species on multi-generational timescales (Geisz et al. 2008; Desforges et al. 2018). In conjunction with increased temperature and drought, the chemical release may impede the growth native flora, but facilitate later recolonization by opportunistic invasive species with resilient seed banks (Bogardi et al. 2012; Nadal et al. 2015). The long-term impacts on faunal species like Salmon or Golden Eagles from physical and chemical damage to critical habitat and food sources have already been substantial, limiting their ability to reproduce or migrate to breeding areas (Kelly 2016; Nawab et al. 2018).

**Next steps: managing system stability**

The increasing interconnectedness and co-location of the developed human and natural world have been a global issue of ongoing concern for generations. Stories as diverse as the Cuyahoga River catching on fire in the 1950s, to the death of predatory birds from DDT in the 1970s, have permeated the literature (Carson 1962; Adler 2014). As human populations continue to grow, this overlap poses increasing challenges to the integrity of natural systems. The warming climate has diminished precipitation and increased regional drought conditions in many Mediterranean ecosystems, increasing the risk of wildfire worsened by invasive species' expansion and fuel loading. After damaging wildfires, chemicals from the built environment can become uncontrolled toxins in the background, further diminishing ecosystem stability and structure.

This cycle of cascading system impacts from the combination of human-driven climate changes and ecosystem encroachment will continue without the proper management of human chemicals. Physical and ecological actions, including invasive species management and runoff entrapment, may slow chemical transport into the environment, providing a first-order defense. This is especially critical as ecosystems attempt to adjust to a new climate regime. Modeling and data software that track the movement and flow of toxins through the environment could be developed to provide forecasts similar to weather reports, allowing for mitigation. A tool to track spatially where there is an overlap of invasive species, subsequent fire range, and potential toxin release would enable resource agencies to improve mitigation and data-collection in high-risk areas. Public education in high-risk neighborhoods could encourage participation in invasive management, minimizing the use of toxic chemicals, and maintaining healthy fire-adapted ecosystems. Alongside public education, however, policy changes to mitigate the negative impacts of combusting building materials and household goods are necessary. As long-lasting, destructive fires expand globally, efforts to plan for fire and toxin mitigation could be widely applied. In the 2018-2019 fire seasons alone, damaging fires in Alaska, California, Australia, Brazil's Amazon, Portugal and, Siberia have destroyed vegetation, critical animal habitat, soil productivity, and countless structures. As climate warming and invasive range expansion continue, the impact of fires, and the toxins they release, will remain.

The need for increased separation between the natural and human systems is clear, with the ongoing development of human communities, creating a significant strain on the natural systems that support them. It is critical to understand the cycle of invasive species expansion, ecosystem disruption, and destabilizing chemical release as a managed system within the global focus on a changing climate. Humans are reliant on ecosystem stability, and increasingly, ecosystems rely on humans.

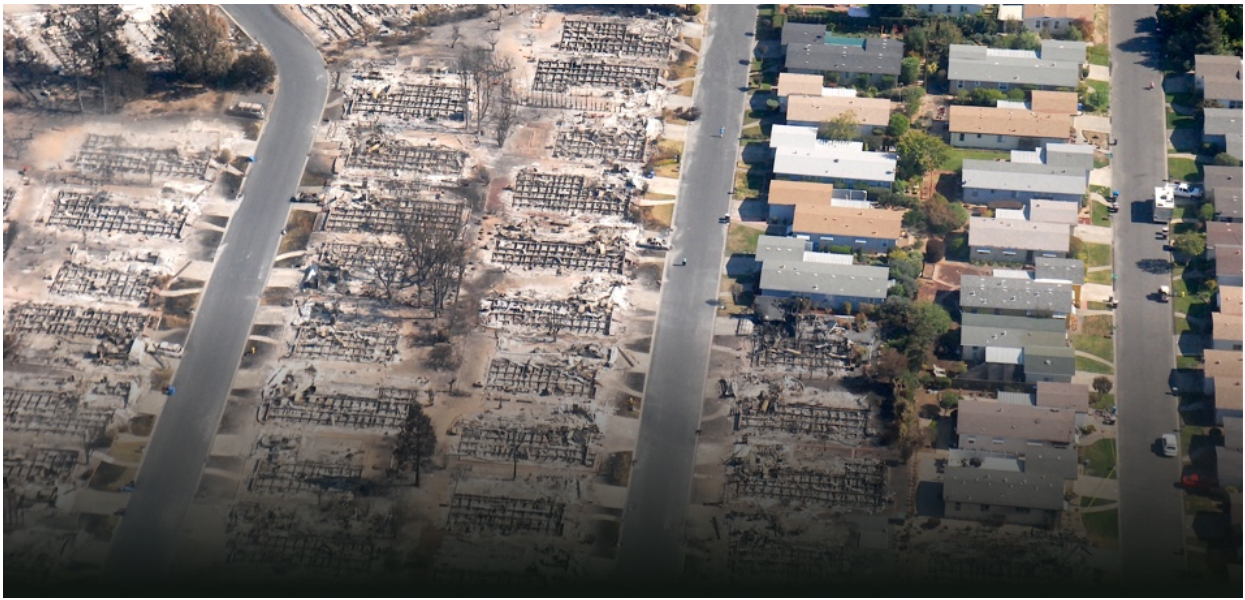

Bibliography
Adler J (2014) The Fable of the Burning River, 45 Years Later. Washington Post

Asner GP, Brodrick PG, Anderson CB, et al. (2016) Progressive forest canopy water loss during the 2012–2015 California drought. Proc Natl Acad Sci 113:E249–E255. DOI: 10.1073/pnas.1523397113

Bell CE, DiTomaso JM, Brooks ML (2017) Invasive Plants and Wildfires

Belmecheri S, Babst F, Wahl ER, et al. (2016) Multi-century evaluation of Sierra Nevada snowpack. Nat Clim Chang. DOI: 10.1038/nclimate2809

Bentz BJ, Régnière J, Fettig CJ, et al. (2010) Climate Change and Bark Beetles of the Western United States and Canada: Direct and Indirect Effects. Bioscience 60:602–613. DOI: 10.1525/bio.2010.60.8.6

Bizzotto EC, Villa S, Vighi M (2009) POP bioaccumulation in macroinvertebrates of alpine freshwater systems. Environ Pollut 157:3192–3198. DOI: http://dx.doi.org/10.1016/j.envpol.2009.06.001

Blais JM, Wilhelm F, Kidd KA, et al. (2003) Concentrations of organochlorine pesticides and polychlorinated biphenyls in amphipods (Gammarus lacustris) along an elevation gradient in mountain lakes of western Canada. Environ Toxicol Chem 22:2605–2613. DOI: 10.1897/02-389

Bogardi JJ, Dudgeon D, Lawford R, et al. (2012) Water security for a planet under pressure: Interconnected challenges of a changing world call for sustainable solutions. Curr Opin Environ Sustain 4:35–43. DOI: 10.1016/j.cosust.2011.12.002

Bogdal C, Naef M, Schmid P, et al. (2009) Unexplained gonad alterations in whitefish (Coregonus spp.) from Lake Thun, Switzerland: Levels of persistent organic pollutants in different morphs. Chemosphere 74:434–440. DOI: 10.1016/j.chemosphere.2008.09.058

California Native Plant Society (2010) Native plants and fire safety Special Issue. Sacramento, CA

Calvino-Cancela M, Lorenzo P, González L, Calviño-Cancela M (2017) Fire increases Eucalyptus globulus seedling recruitment in forested habitats: Effects of litter, shade and burnt soil on seedling emergence and survival. Artic For Ecol Manag. DOI: 10.1016/j.foreco.2017.12.018

Carson R (1962) Silent Spring

Collins BM, Miller JD, Knapp EE, Sapsis DB (2019) A quantitative comparison of forest fires in central and northern California under early (1911–1924) and contemporary (2002–2015) fire suppression. Int J Wildl Fire 28:138. DOI: 10.1071/wf18137

Daly GL, Wania F (2005) Critical Review Organic Contaminants in Mountains. Environ Sci Technol 39:385–398. DOI: 10.1021/es048859u

Denevan WM (1992) The Pristine Myth: The Landscape of the Americas in 1492. Assoc Am Geogr 82:369–385

Denise K. Reaves, Erika Ginsburg, John J. Bang, and JMF (2015) Persistent organic pollutants and obesity: potential mechanisms for breast cancer promotion? Endocr Relat cancer 28:25–38. DOI: 10.1530/ERC-14-0411.Persistent

Desforges JP, Hall A, McConnell B, et al. (2018) Predicting global killer whale population collapse from PCB pollution. Science (80- ) 361:1373–1376.DOI: 10.1126/science.aat1953

Diffenbaugh NS, Swain DL, Touma D (2015) Anthropogenic warming has increased drought risk in California. Proc Natl Acad Sci 112:3931–3936.DOI 10.1073/pnas.1422385112

Elliott JE, Levac J, Guigueno MF, et al. l.l (2012) Factors influencing legacy pollutant accumulation in alpine osprey: Biology, topography, or melting glaciers? Environ Sci Technol 46:9681–9689.DOI: 10.1021/es301539b

EPA U (2016) Wildfire Smoke: A Guide for Public Health Officials

Fernández-Fernández M, Gómez-Rey MX, González-Prieto SJ (2015) Effects of fire and three fire-fighting chemicals on main soil properties, plant nutrient content and vegetation growth and cover after 10 years. Sci Total Environ 515–516:92–100. doi: 10.1016/j.scitotenv.2015.02.048

Fire C (2017) News release: Record 129 Million Dead Trees in California. Sacramento, CA

Flato G, Marotzke J, Abiodun B, et al. (2013) Evaluation of Climate Models. Clim Chang 2013 Phys Sci Basis Contrib Work Gr I to Fifth Assess Rep Intergov Panel Clim Chang 741–866. DOI: 10.1017/CBO9781107415324

Friedman CL, Selin NE (2016) PCBs in the Arctic atmosphere: determining important driving forces using a global atmospheric transport model. Atmos Chem Phys 16:3433–3448. DOI: 10.5194/acpd-15-30857-2015

Geisz HN, Dickhut RM, Cochran MA, et al. (2008) Melting glaciers: A probable source of DDT to the Antarctic marine ecosystem. Environ Sci Technol 42:3958–3962. DOI: 10.1021/es702919n

Hansen KM, Christensen JH, Geels C, et al. (2015) Modelling the impact of climate change on atmospheric transport and the fate of persistent organic pollutants in the Arctic. Atmos Chem Phys 15:6549–6559. DOI: 10.5194/acp-15-6549-2015

He T, Belcher CM, Lamont BB, Lim SL (2015) A 350-million-year legacy of fire adaptation among conifers. DOI: 10.1111/1365-2745.12513

IPCC Working Group 1 (2014) IPCC Fifth Assessment Report (AR5) - The physical science basis. IPCC

Johnston A (2018) After the fires, rain- and the threat of toxic runoff. KALW San Fr. 1–9

Keeley JE (2002) Native American impacts on fire regimes of the California coastal ranges. J Biogeogr 29:303–320

Keeley JE (2001) Fire and invasive species in Mediterranean-climate ecosystems of California. Pages 81-94 in. Miscellaneous Publication

Kelly EC (2016) The Listing of Coast Redwood as Endangered Under the IUCN Red List: Lessons for Conservation 1

Kibet J, Bosire J, Kinyanjui T (2017) Characterization of Forest Fire Emissions and Their Possible Toxicological Impacts on Human Health. J For Environ Sci 33:113–121. DOI: 10.7747/JFES.2017.33.2.113

Kilgore BM, Taylor D (1979) Fire History of a Sequoia-Mixed Conifer

Kloos J, Gebert N, Rosenfeld T, Renaud FG (2013) Climate change, water conflicts, and human security: regional assessment and policy guidelines for the Mediterranean, Middle East, and the Sahel. 256

Knelman JE, Graham EB, Trahan NA, et al. (2015) Fire severity shapes plant colonization effects on bacterial community structure, microbial biomass, and soil enzyme activity in secondary succession of a burned forest. Soil Biol Biochem 90:161–168. DOI: 10.1016/j.soilbio.2015.08.004

Lafrenière MJ, Blais JM, Sharp MJ, Schindler DW (2006) Organochlorine pesticide and polychlorinated biphenyl concentrations in snow, snowmelt, and runoff at Bow Lake, Alberta. Environ Sci Technol 40:4909–4915. DOI: 10.1021/es060237g

Lee Y-M, Kim K-S, Kim S-A, et al. (2014) Prospective associations between persistent organic pollutants and metabolic syndrome: A nested case-control study. Sci Total Environ 496:219–25. DOI: 10.1016/j.scitotenv.2014.07.039

Little JB (2018) The California Indigenous Peoples Using Fire for Agroforestry. Pacific Stand.

Mackay D, Celsie AKD, Powell DE, Parnis JM (2018) Bioconcentration, bioaccumulation, biomagnification, and trophic magnification: a modeling perspective. Environ Sci Process Impacts 20:72–85. DOI: 10.1039/C7EM00485K

Miner KR, Blais J, Bogdal C, et al. (2017) Legacy organochlorine pollutants in glacial watersheds: A review. Environ Sci Process Impacts 19:1–10. DOI: 10.1039/c7em00393e

Miner KR, Campbell S, Gerbi C, et al. (2018a) Organochlorine pollutants within a polythermal glacier in the Interior Eastern Alaska Range. Water 10:1–14. DOI: 10.3390/w10091157

Miner KR, Kreutz KJ, Jain S, et al. (2018b) A screening-level approach to quantifying risk from the glacial release of organochlorine pollutants in the Alaskan Arctic. J. Expo. Sci. Environ. Epidemiol.

Miner KR, Kreutz KJ, Jain S, et al. (2018c) A screening-level approach to quantifying risk from the glacial release of organochlorine pollutants in the Alaskan Arctic. J Expo Sci Environ Epidemiol. DOI: doi.org/10.1038/s41370-018-0100-7

Nadal M, Marquès M, Mari M, Domingo JL (2015) Climate change and environmental concentrations of POPs: A review. Environ Res 143:177–185. DOI: 10.1016/j.envres.2015.10.012

Nawab J, Khan S, Xiaoping W (2018) Ecological and health risk assessment of potentially toxic elements in the major rivers of Pakistan: General population vs. Fishermen. Chemosphere 202:154–164. DOI: 10.1016/j.chemosphere.2018.03.082

Ngwa EN, Kengne A-P, Tiedeu-Atogho B, et al. (2015) Persistent organic pollutants as risk factors for type 2 diabetes. Diabetol Metab Syndr 7:41. DOI: 10.1186/s13098-015-0031-6

Noestheden M, Dennis EG, Zandberg WF (2018) Quantitating Volatile Phenols in Cabernet Franc Berries and Wine after On-Vine Exposure to Smoke from a Simulated Forest Fire. J Agric Food Chem 66:695–703. DOI: 10.1021/acs.jafc.7b04946

Parsons DJ, Debenedetti SH (1979) Impact of Fire Suppression on a Mixed-Conifer Forest. For Ecol Manage 2:21–33

Rao LE, Allen EB (2010) Combined effects of precipitation and nitrogen deposition on native and invasive winter annual production in California deserts. Oecologia 162:1035–1046. DOI: 10.1007/s00442-009-1516-5

Reid C, Brauer M, Johnston F, et al. (2016) Critical Review of Health Impacts of Wildfire Smoke Exposure. Environ Health Perspect 124:1334–1343. DOI: 10.1289/ehp.1409277

Ren J, Wang X, Wang C, et al. (2016) Biomagnification of persistent organic pollutants along a high-altitude aquatic food chain in the Tibetan Plateau: Processes and mechanisms. Environ Pollut 1–8. DOI: 10.1016/j.envpol.2016.10.019

Revelle R, Suess HE (1957) Carbon Dioxide Exchange Between Atmosphere and Ocean and the Question of an Increase of Atmospheric CO2 during the Past Decades Carbon Dioxide Exchange Between Atmosphere and Ocean and the Question of an Increase of Atmospheric CO, during the Past Decades. Tellus 9: DOI: 10.3402/tellusa.v9i1.9075

Roe GH, Baker MB, Herla F (2016) Centennial glacier retreat as categorical evidence of regional climate change. Nat Geosci 1:. DOI: 10.1038/ngeo2863

Ruzzin J (2012) Public health concern behind the exposure to persistent organic pollutants and

the risk of metabolic diseases. BMC Public Health 12:298. DOI: 10.1186/1471-2458-12-298

Schwartz J (2018) More floods and more droughts: Climate change delivers both. New York Times

Steel ZL, Safford HD, Viers JH (2015) The fire frequency-severity relationship and the legacy of fire suppression in California forests. Ecosphere 6:art8. DOI: 10.1890/es14-00224.1

Treut L, Somerville R, Cubasch U, et al. (2007) Historical Overview of Climate Change Science

Villa S, Migliorati S, Monti GS, et al. (2017) Risk of POP mixtures on the Arctic food chain. Environ Toxicol Chem 36:1181–1192. DOI: 10.1002/etc.3671

Walters DM, Jardine TD, Cade BS, et al. (2016) Trophic Magnification of Organic Chemicals: A Global Synthesis. Environ Sci Technol 50:4650–4658. DOI: 10.1021/acs.est.6b00201

Wang X, Gong P, Wang C, et al. (2016) A review of current knowledge and future prospects regarding persistent organic pollutants over the Tibetan Plateau. Sci Total Environ 573:139–154. DOI: 10.1016/j.scitotenv.2016.08.107